\documentclass[a4paper,referee]{raa}            

\usepackage{graphicx, times}             
\usepackage{natbib}
\usepackage{amssymb,amsmath}
\usepackage{xcolor} 
\bibpunct{(}{)}{;}{a}{}{,}

\usepackage[a4paper=true,pagebackref=true]{hyperref}
\hypersetup{pdftitle = The title of my PDF, pdfauthor = My name, pdfsubject= The subject, pdfkeywords = keyword1 keyword2 keyword3} 
\hypersetup{colorlinks = true, linkcolor = green, anchorcolor = red, citecolor = blue, filecolor = red, pagecolor = red, urlcolor = red}

\begin{document}

  \title{The Mini-SiTian Array: Design and application of Master Control System}

   \volnopage{Vol.0 (20xx) No.0, 000--000}      
   \setcounter{page}{1}          
   
    \author{Zheng Wang\inst{1}, Jin-Hang Zou\inst{1}, Liang Ge\inst{1,2}, Min He\inst{1}, Jian Li\inst{1}, Yi Hu\inst{1}, Jian-Feng Tian\inst{1}}

   \institute{National Astronomical Observatories, Chinese Academy of Sciences,
             Beijing 100010, China;\\
              \email{wzheng@bao.ac.cn}
         \and
             Institute for Frontiers in Astronomy and Astrophysics, Beijing Normal University, Beijing 100875, China\\
             }
             

\vs\no
   {\small Received 20xx month day; accepted 20xx month day}

\abstract{ The SiTian Project represents a groundbreaking initiative in astronomy, aiming to deploy a global network of telescopes, each with a 1-meter aperture, for comprehensive time-domain sky surveys. The network's innovative architecture features multiple observational nodes, each comprising three strategically aligned telescopes equipped with filters. This design enables three-color (g, r, i) channel imaging within each node, facilitating precise and coordinated observations. As a pathfinder to the full-scale project, the Mini-SiTian Project serves as the scientific and technological validation platform, utilizing three 30-centimeter aperture telescopes to validate the methodologies and technologies planned for the broader SiTian network.
This paper focuses on the development and implementation of the Master Control System (MCS),and the central command hub for the Mini-SiTian array. The MCS is designed to facilitate seamless communication with the SiTian Brain, the project's central processing and decision-making unit, while ensuring accurate task allocation, real-time status monitoring, and optimized observational workflows. The system adopts a robust architecture that separates front-end and back-end functionalities.
A key innovation of the MCS is its ability to dynamically adjust observation plans in response to transient source alerts, enabling rapid and coordinated scans of target sky regions. The paper provides an in-depth analysis of the system's internal components, including the communication system, which is critical for seamless network operation. Extensive testing has validated the functionality, reliability, and compatibility of these components within the overall system architecture.
The successful deployment of the MCS in managing the Mini-SiTian array has demonstrated its practicality and efficacy in collaborative observation and distributed control. By simplifying cluster management and ensuring data integrity, the MCS represents a significant advancement in astronomical observation control systems. Its scalable and adaptable design not only supports the future expansion of the SiTian network but also provides a blueprint for other large-scale telescope arrays, marking a transformative step forward in time-domain astronomy.
}

   \authorrunning{Zheng Wang }            
   \titlerunning{The Mini-SiTian Array: Design and application of Master Control System}  

   \maketitle

%
\section{Introduction}           
\label{sect:intro}
The SiTian Project is a next-generation time-domain survey,aiming to establish a globally distributed network of about 60 1-meter-class telescopes for continuous sky monitoring. For such a large-scale telescope network, the observation control system must meet exceptionally high demands in automated control, cluster deployment, and robust maintenance. However, existing control systems, such as the Remote Telescope System 2 (RTS2) and the Universal Observation Control System (UOCS), while reliable for small to medium-sized telescopes, face significant limitations in scalability, rapid response, and large-scale coordination \cite{peter2006,wang2021}.

RTS2's inheritance-based architecture struggles with the complexity of managing extensive arrays, leading to inefficiencies in large-scale operations. Similarly, UOCS, despite its precision in the Large Sky Area Multi-Object Fiber Spectroscopic Telescope (LAMOST) control \cite{cui2012,peter2008}, lacks the agility required for real-time transient event detection and rapid task reconfiguration. These limitations highlight the critical need for a next-generation control system specifically designed to address the unique demands of the SiTian Project, ensuring seamless coordination, scalability, and responsiveness across a global telescope network.

To address these limitations, we developed the Sitian Observation Control System (S-OCS), a control system integrating UOCS's precise communication protocols with RTS2's robust state management. This hybrid architecture enhances agility and efficiency, enabling rapid adjustments for large-scale arrays. The Mini-SiTian Project, featuring three prototype telescopes of 30cm aperture, serves as the technological testbed for validating S-OCS. Rigorous long-term stability testing on this platform has demonstrated the system's reliability and readiness for full-scale deployment.

While the S-OCS was initially designed for single-telescope autonomous control, we developed the Master Control System (MCS) to coordinate the Mini-SiTian array. The MCS serves as the central hub for autonomous operation, enabling seamless communication with the SiTian Brain, optimal task allocation, and real-time telescope status monitoring. These functionalities streamline data flow, reduce computational load, and enhance system efficiency. Additionally, the MCS incorporates robust security measures, including an isolated control environment, to safeguard against external interference and ensure data integrity. Through its advanced control mechanisms, the MCS establishes a new benchmark for large-scale astronomical observation systems.

As the central component of the Mini-SiTian array's information infrastructure, the MCS guarantees efficient data transmission and rapid response to transient events. Through its seamless integration with the SiTian Brain and advanced control mechanisms, the MCS plays a pivotal role in advancing the scientific objectives of the SiTian Project. This innovative system not only addresses the limitations of existing control frameworks but also sets a new standard for large-scale astronomical observation, paving the way for groundbreaking discoveries in time-domain astronomy.


\section{System Functions and Design}
\label{sect:Obs}
The MCS is a cornerstone of the SiTian Project's observation control infrastructure, serving as the central hub that integrates and coordinates the SiTian Brain and the S-OCS. As illustrated in Figure 1, the MCS is strategically positioned at the intersection of these two systems, enabling seamless communication and operational synchronization to achieve the project's scientific objectives.

As an intermediate exchange hub, the MCS will implement the following functions:
\begin{enumerate}
\item Observation Plan Reception and Parsing: The MCS receives comprehensive sky survey observation plans from the SiTian Brain, which are meticulously parsed into actionable directives. These plans are then translated into specific control commands tailored to the operational parameters of each telescope within the array. This process ensures precise coordination and consistency across all telescopes, aligning their activities with the project's overarching observational strategy.

\item Command Distribution and Execution: The MCS employs an optimized command distribution mechanism to deliver control instructions to individual telescopes. This mechanism ensures minimal latency and high reliability, critical for maintaining synchronized operations across the array. The system's ability to dynamically adjust commands in response to real-time conditions further enhances its operational efficiency.

\item Telescope Status Monitoring and Feedback: The MCS continuously monitors the operational status of all telescopes, collecting data on positional accuracy, environmental conditions, and instrumental health. This information is analyzed in real time and fed back to the SiTian Brain, enabling performance evaluation and adaptive adjustments. The feedback loop is essential for maintaining system integrity and addressing deviations from optimal performance promptly.
\end{enumerate}
To support its multifaceted roles with robustness and efficiency, the MCS architecture is divided into front-end and back-end modules:
\begin{enumerate}
\item Front-End Module: Provides an intuitive user interface for operators to monitor system status, adjust parameters, and initiate manual overrides when necessary. The interface is designed for ease of use while ensuring comprehensive access to critical system functions.
\item Back-End Module: Handles complex computational tasks, including plan parsing, command generation, and data analysis. Advanced algorithms ensure high precision and speed, enabling the MCS to manage large-scale operations seamlessly.
\end{enumerate}

This structural design ensures that the MCS is engineered for maximum reliability, incorporating built-in redundancies and error-handling mechanisms to guarantee uninterrupted operation. Efficient interaction between the MCS, SiTian Brain, and S-OCS is achieved through optimized communication protocols and data exchange formats, minimizing latency and maximizing throughput. These design features are critical for maintaining the stability of the Mini-SiTian array and ensuring the quality of observational data.

As a pivotal component of the SiTian Project, the MCS enables the precise execution of its ambitious observational goals. Through its advanced capabilities in plan reception, command distribution, real-time monitoring, and feedback, the MCS sets a new standard for large-scale astronomical observation control systems. Its scalable and adaptable architecture not only supports the current Mini-SiTian array but also provides a robust foundation for the future expansion of the SiTian network and other large-scale telescope arrays.

   \begin{figure}
   \centering
   \includegraphics[width=\textwidth, angle=0]{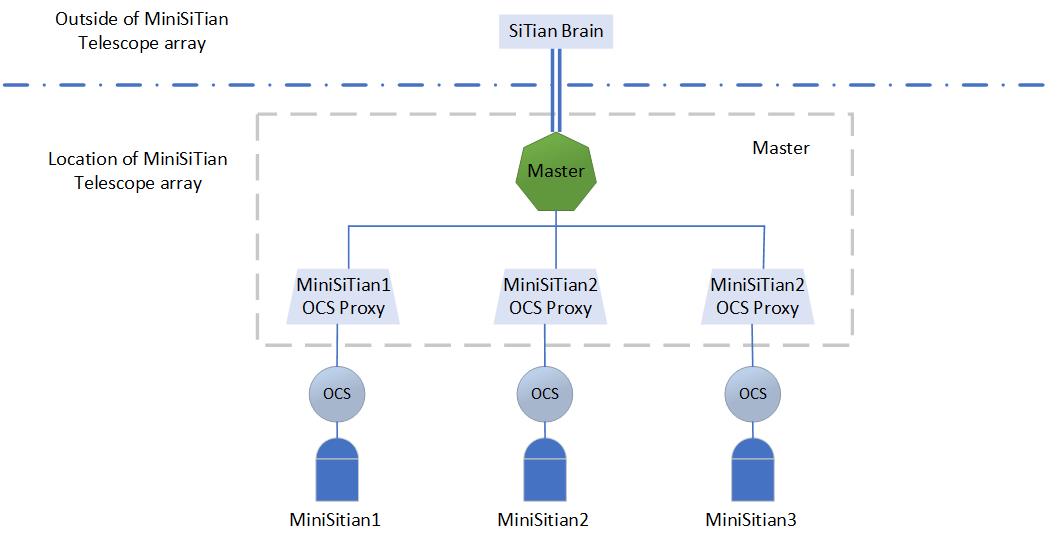}
   \caption{ The MCS Architecture Diagram }
   \label{Figure1}
   \end{figure}

\section{ System Implementation}
\label{sect:Implementation}

The MCS is implemented through a meticulously designed modular architecture, where each module is assigned specific business functions. These modules operate both independently and collaboratively, ensuring seamless system operation while preventing mutual interference. This distributed control approach enhances the system's scalability and adaptability, making it well-suited for large-scale astronomical observation tasks. As illustrated in Figure 2, the MCS comprises the following core modules: the SiTian Brain Proxy Module, the Communication System Module, the Information Display Module, the Observation Plan Management Module, the Observation Mode Control Module, the Status Collection Module, and the Observation Control System Proxy (OCS Proxy) Module. Each module is optimized for its designated role, collectively forming the foundation of the MCS's robust functionality.

   \begin{figure}
   \centering
   \includegraphics[width=\textwidth, angle=0]{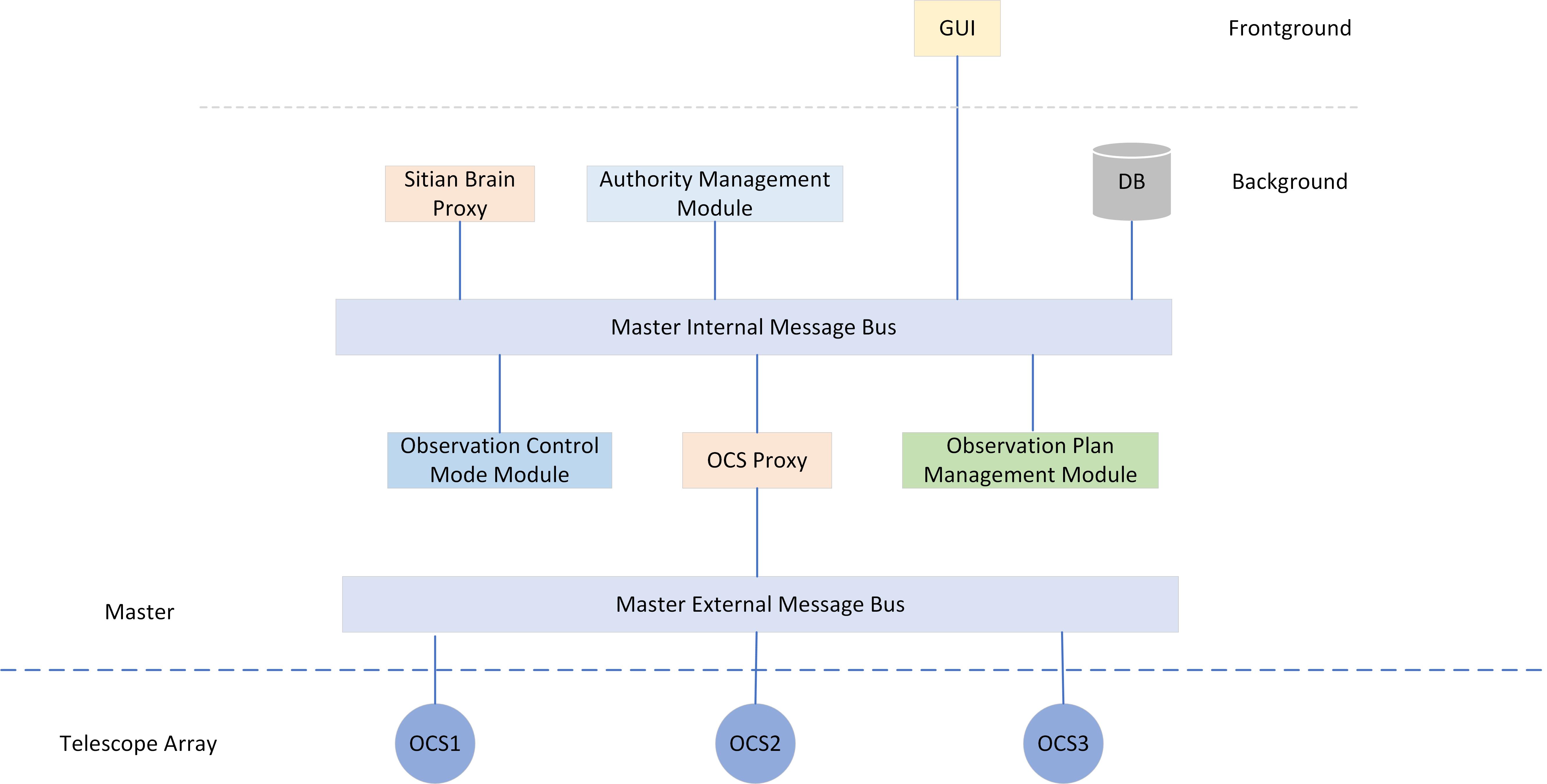}
   \caption{ Internal Structure of the MCS}
   \label{Figure2}
   \end{figure}

\subsection{System Development and Operating Environment}
The MCS is developed on the Windows 10 operating system, with Python 3.9 serving as the primary programming environment. Through rigorous scientific research and detailed engineering requirement analysis, we have carefully selected several Python third-party libraries to enhance the system's efficiency, stability, and performance. These libraries are critical for ensuring the MCS's high-performance operation and robust functionality.
\begin{enumerate}
    \item ZeroMQ (ZMQ) is an advanced communication library specifically designed for distributed control systems[ \cite{dworak2010}]. It provides a lightweight, asynchronous message-passing mechanism that supports multiple communication patterns, including request-response, publish-subscribe, and push-pull. These capabilities establish a robust foundation for inter-module communication within the MCS, enabling efficient and reliable data exchange across the system.
    \item pySide2 is a Python binding for the Qt framework, enabling the development of cross-platform graphical user interfaces (GUIs). Its extensive functionality and high flexibility have significantly enhanced the MCS's user interface design and interaction. The resulting intuitive and user-friendly interface improves operational efficiency and overall user experience.
    \item Asyncio, introduced in Python 3.4 as part of the standard library, enables concurrent handling of multiple operations within a single process. The MCS leverages asyncio to implement asynchronous control of multiple tasks and devices. This approach not only improves the system's response time and operational efficiency but also simplifies the management of telescope control commands, thereby enhancing the system's robustness and scalability.
\end{enumerate}
The integration of these third-party libraries not only enhances development efficiency and reduces system complexity but also enables distributed communication between modules and concurrent execution of multiple tasks. This combination of technologies ensures that the MCS meets the demanding requirements of large-scale astronomical observation systems.

\subsection{Front-End and Back-End Separation}
This architectural approach strictly distinguishes between business logic and user interface display, significantly enhancing the system's stability, maintainability, and scalability. By isolating core computational tasks from user interaction, the back-end module focuses on efficient data processing, command generation, and system management. This separation not only facilitates independent development and optimization of each component but also ensures the system's adaptability to future requirements, enabling seamless evolution and long-term sustainability.
\subsubsection{Front-End: Information Display and Human-Computer Interaction}
The front-end of the MCS is designed to provide an intuitive and user-friendly interface for system information display and human-computer interaction. It enables observation personnel to efficiently manage observation tasks, monitor the real-time status of array telescopes, and access critical operational data. The module prioritizes a seamless and interactive user experience, ensuring that personnel can quickly respond to system status changes and make informed decisions. Key functionalities of the front-end include:
\begin{enumerate}
    \item Task Management Interface: Enables users to create, edit, and manage observation tasks with ease, ensuring efficient task allocation and execution.
    \item Real-Time Status Dashboard: Offers a comprehensive overview of the operational status of all telescopes within the array, including positional accuracy, environmental conditions, and instrumental health.
    \item Alert and Notification System: Provides timely notifications for significant events or system status changes, allowing users to take prompt and appropriate actions.
\end{enumerate}

By integrating these features, the front-end module enhances operational efficiency and user experience, ensuring that the MCS meets the demanding requirements of information display and array operation control.
\subsubsection{Back-End: Business Logic Processing}
The back-end of the MCS serves as the core engine for business processing, executing critical system operations and managing complex workflows. It is responsible for implementing specific business logic, processing instructions and requests from the front-end, performing data processing and logical calculations, and transmitting results back to the front-end for display. The back-end module plays a pivotal role in key functionalities, including observation plan management, telescope control, and system status monitoring. Its primary responsibilities include:
\begin{enumerate}
    \item Observation Plan Management: Handles the formulation, scheduling, and execution of observation plans, ensuring alignment with scientific objectives and operational constraints.
    \item Telescope Control: Manages control commands for individual telescopes, enabling precise and coordinated observations across the array. This includes real-time adjustments based on environmental conditions and target priorities.
    \item System Status Monitoring: Continuously monitors the health and performance of the system, collecting data on telescope status, environmental conditions, and instrumental health. This information is analyzed and fed back to the front-end in real time, enabling adaptive decision-making.
\end{enumerate}

The separation of the front-end and back-end ensures that the MCS can efficiently handle complex business logic while maintaining a responsive and user-friendly interface. This modular design not only enhances system stability and maintainability but also facilitates future enhancements and scalability, ensuring the system's adaptability to evolving scientific and operational requirements.

\subsection{Communication System Module}
At the core of its communication infrastructure, the MCS employs the publish-subscribe (PUB-SUB) model of ZMQ as the primary message bus. This mechanism allows each module to publish or subscribe to messages based on its specific requirements, while the message bus ensures accurate and reliable message forwarding. By eliminating the need for direct inter-module communication, this design enables each module to focus on its core business logic, significantly improving the system's scalability and adaptability.

To meet the system's real-time and functional requirements, messages are categorized into three main types:
\begin{enumerate}
    \item Event Messages: Event messages are notification messages generated by the MCS following data processing. They inform other modules within the system to perform subsequent operations on the processed data. These messages have stringent real-time requirements, ensuring rapid system response and execution of follow-up tasks. Examples include notifications of completed observations, data readiness for analysis, and system state changes.
    \item Command Control Messages: Command control messages are critical in non-automatic observation modes. The MCS uses these messages to generate and transmit control commands directly to the telescopes, enabling precise operational control. This functionality supports specific observation tasks while ensuring efficient system performance. Examples include initiating observations, adjusting telescope configurations, and managing observational sequences.
    \item Status Messages: Status messages provide real-time operating status information for each telescope in the array. They offer a comprehensive operational overview, enabling the MCS to monitor and adjust telescope statuses promptly. These messages are essential for maintaining system stability by delivering continuous feedback on telescope health and performance. Examples include telescope position, environmental conditions, and instrumental health data, facilitating proactive system management.
\end{enumerate}
This modular and message-driven architecture ensures that the MCS can efficiently handle complex operational workflows while maintaining high reliability and adaptability, making it well-suited for large-scale astronomical observation systems.

\subsection{SiTian Brain Proxy Module}

During the operational cycle of the telescopes, each node continuously transmits real-time operational data to the SiTian Brain. Based on predefined scientific objectives and the actual operating conditions of each telescope, the SiTian Brain plans and schedules observation tasks. In the case of high-priority transient source events requiring rapid response, the SiTian Brain dynamically selects the most suitable node to execute observation tasks. When no specific observation plan is issued by the SiTian Brain, each node independently formulates and executes its observation plan to ensure continuous and efficient operation.
The SiTian Brain Proxy Module plays a critical role in ensuring secure and stable communication between the MCS and the SiTian Brain. It is responsible for encrypting all communication data and parsing commands, thereby maintaining system controllability and security. As illustrated in Figure 3, the proxy module establishes communication with the SiTian Brain using the TCP/IP protocol and implements RSA security authentication during the initial connection. This ensures that only authenticated users can transmit data, effectively preventing unauthorized access and data leakage.

   \begin{figure}
   \centering
   \includegraphics[width=\textwidth, angle=0]{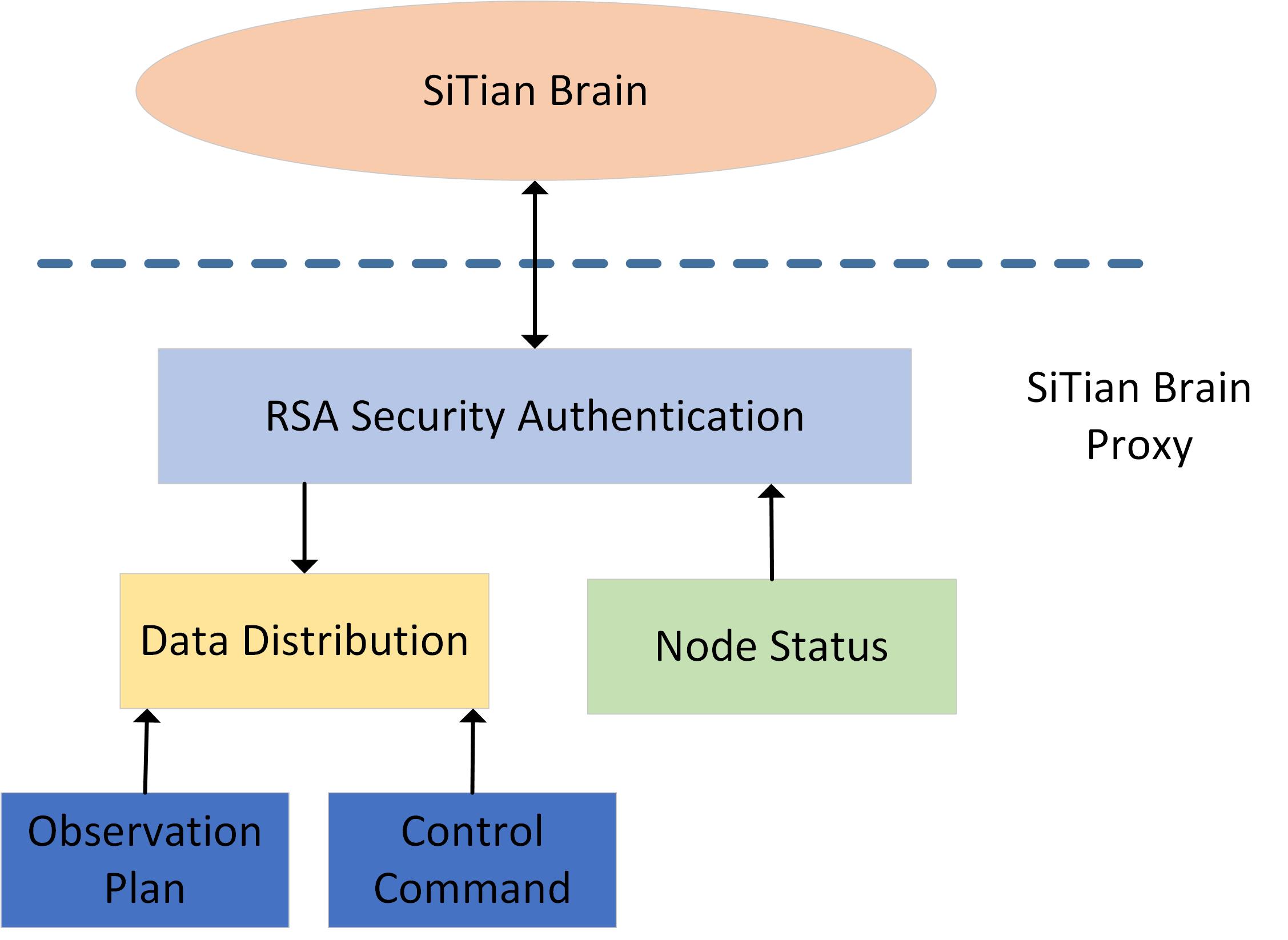}
   \caption{ SiTian Brain Proxy Structure}
   \label{Figure4}
   \end{figure}

The data processed by the proxy module are mainly divided into two major categories: observation plans and control commands. Observation plans may include regular multi-target observations or specific target observations for high-priority transient sources. Control commands target specific equipment within the node, such as instructing the Mini-Sitian1 camera which is the telescope of Mini-SiTian array to take two exposures, each lasting 10 seconds.

Additionally, the proxy module integrates and transmits the operational status of the telescopes, meteorological conditions, and other relevant information to the SiTian Brain. This enables real-time monitoring of all equipment within the node, facilitating timely adjustments and optimizations to maintain system performance and data quality.

\subsection{ Observation Control Mode Management Module}
The SiTian Brain does not directly control telescope equipment during observation tasks. Instead, it formulates detailed observation plans based on scientific requirements and transmits these plans to the MCS for execution. To accommodate diverse operational needs, the MCS provides five observation control modes, each tailored to specific control permissions and task requirements:
\begin{enumerate}
    \item Fully Automatic Observation Control Mode: In this mode, the MCS autonomously identifies observation targets and coordinates the activities of all telescopes without manual intervention. This mode ensures efficient and autonomous task execution, maximizing system throughput while minimizing human error. It is ideal for routine observations and large-scale surveys.
    \item Semi-Automatic Observation Control Mode: The MCS distributes observation plans to individual telescopes, which then execute tasks automatically. The MCS monitors the operational status of each telescope, enabling real-time adjustments and interventions when necessary. This mode balances automation with manual oversight, making it suitable for complex observation scenarios.
    \item Manual Observation Control Mode: In this mode, observation personnel manually select observation targets using Right Ascension and Declination coordinates and designate specific telescopes via the MCS. Staff can define custom observation tasks for each telescope, significantly enhancing flexibility and enabling targeted observations or specialized projects that require precise human oversight.
    \item Debugging Control Mode: Designed for equipment testing and maintenance, this mode allows observation personnel to send individual control commands to various telescope components. It provides the flexibility and control needed for system troubleshooting, ensuring rapid identification and resolution of issues.
    \item Information Display Mode: In this mode, the MCS focuses solely on monitoring telescope status without issuing control commands. It provides a comprehensive overview of system operations, facilitating routine monitoring and data analysis. This mode is particularly useful for real-time status surveillance and data visualization.
\end{enumerate}
The diversity of these five modes provides the MCS with a wide range of operational capabilities. In the semi-automatic mode, control rights are delegated to the S-OCS, while in other modes, decision-making authority resides with the MCS. This architectural design ensures the system can adapt to various observation requirements and task scenarios, combining the efficiency of automated systems with the flexibility of manual control when needed.

\subsection{OCS Proxy Module}
In the MCS architecture, the OCS Proxy Module serves as a critical intermediary between the SiTian Brain and the telescopes. It is primarily responsible for executing permission authentication and forwarding observation information. In automatic observation mode, the OCS Proxy Module delegates specific control tasks to the S-OCSs, which manage the telescope observation process and equipment operation. This separation of responsibilities ensures efficient task execution and system scalability.

Core Functions of the OCS Proxy Module
The OCS Proxy Module establishes a robust communication channel with each S-OCS, enabling precise coordination and oversight of observation tasks. It receives observation plans, emergency observation targets, and specific control commands from the MCS, accurately relaying these directives to the corresponding S-OCS. Upon task completion, the module initiates a target completion event and notifies other system modules, ensuring seamless task progression.

Beyond task management, the OCS Proxy Module plays a vital role in monitoring the observation process and maintaining system integrity. Its key responsibilities include:
\begin{enumerate}
    \item Task Execution Monitoring: Tracks the execution status of observation tasks to ensure adherence to planned schedules and objectives.
    \item State Feedback: Aggregates and transmits real-time status updates from the telescopes to the MCS, including operational health metrics, environmental conditions, and detected anomalies.
    \item Error Handling: Identifies and reports errors or deviations in the observation process, enabling swift corrective actions to minimize disruptions.
    \item Resource Management: Optimizes the utilization of telescopes and associated equipment, efficiently allocating resources to meet the demands of diverse observation tasks.
\end{enumerate}
By maintaining a robust communication and monitoring framework, the OCS Proxy Module enables the MCS to effectively manage and control telescopes, even in complex and dynamic observational scenarios. This module is indispensable for ensuring the integrity, reliability, and efficiency of the entire observation control system, allowing scientific objectives to be achieved with precision and consistency.

\subsection{Interface Module}
The Interface Module of the MCS, developed using PySide2, serves as the central hub for operator-system interaction and system-wide information aggregation. It provides observation personnel with an intuitive and user-friendly platform to monitor, control, and manage the telescope array. The module's design prioritizes user experience, featuring a logically organized layout and simplified operation processes that minimize learning costs and enhance work efficiency, even for non-technical personnel.
The Interface Module offers the following key functionalities:
\begin{enumerate}
    \item System Status Information Summary: This module offers a comprehensive, real-time monitoring platform for the operational status of telescopes. It displays critical information such as telescope connection status, ongoing observation tasks, and equipment conditions, providing a macro-level overview of the entire observation system.
    \item Operation Control: This module empowers observation personnel with full control over the telescopes. It provides a complete set of control commands, enabling the initiation, cessation, pausing, and adjustment of observation tasks. Additionally, it supports remote control and equipment setting adjustments, ensuring flexible and efficient management of the observation process.
    \item Observation Records: The Interface Module displays daily observation plans and summarizes the observation records of the telescope array. This functionality allows observers to quickly assess observation progress, verify results, and ensure alignment with scientific objectives.
\end{enumerate}
The Interface Module is designed with a focus on accessibility and usability. Its intuitive interface and logically structured layout enable seamless interaction, even for users without technical expertise. By simplifying operation processes and providing clear visual feedback, the module significantly reduces user learning curves and enhances overall operational efficiency.

\section{Testing and Application}

Currently, each telescope within the Mini-SiTian array is equipped with individual automated control capabilities through the S-OCS. Leveraging this functionality, we conducted rigorous testing of the MCS in two critical areas: observation control authority and transient source distribution. These tests were designed to validate the MCS's ability to effectively manage and coordinate the SiTian telescope array. The specific details of these evaluations are outlined below.

\subsection{Observation Control authority Testing}
In the practical operation and control of the telescopes, the S-OCS parses observation targets into a sequence of command control flows. It then dispatches commands to the telescope components based on the observation control script and the status of command execution, ultimately achieving precise tracking and data acquisition of the designated observation target. The observation targets are sourced from two distinct origins, depending on the control permissions: one originates from the MCS, and the other from the S-OCS's intrinsic observation target database. To ensure seamless integration and functionality across all observation modes, we conducted comprehensive testing. The specifics of these tests are detailed below:

\begin{itemize}
    \item Automatic Observation Control Mode: As illustrated in Figure 4, the target 'test1' was selected from the imported observation plan. The MCS initiated the automatic execution of subsequent targets, starting with this designated target. Each telescope's S-OCS within the node array received the same observation target from the MCS and commenced observations under the governance of their respective S-OCS units.
    The S-OCS's role was limited to receiving the observation target and providing feedback to the MCS upon task completion, subsequently requesting the next target. The MCS dispatched a new target only after all telescopes in the node successfully completed the current observation task. This coordinated process ensured seamless and efficient observation across the array.
   \begin{figure}
   \centering
   \includegraphics[width=\textwidth, angle=0]{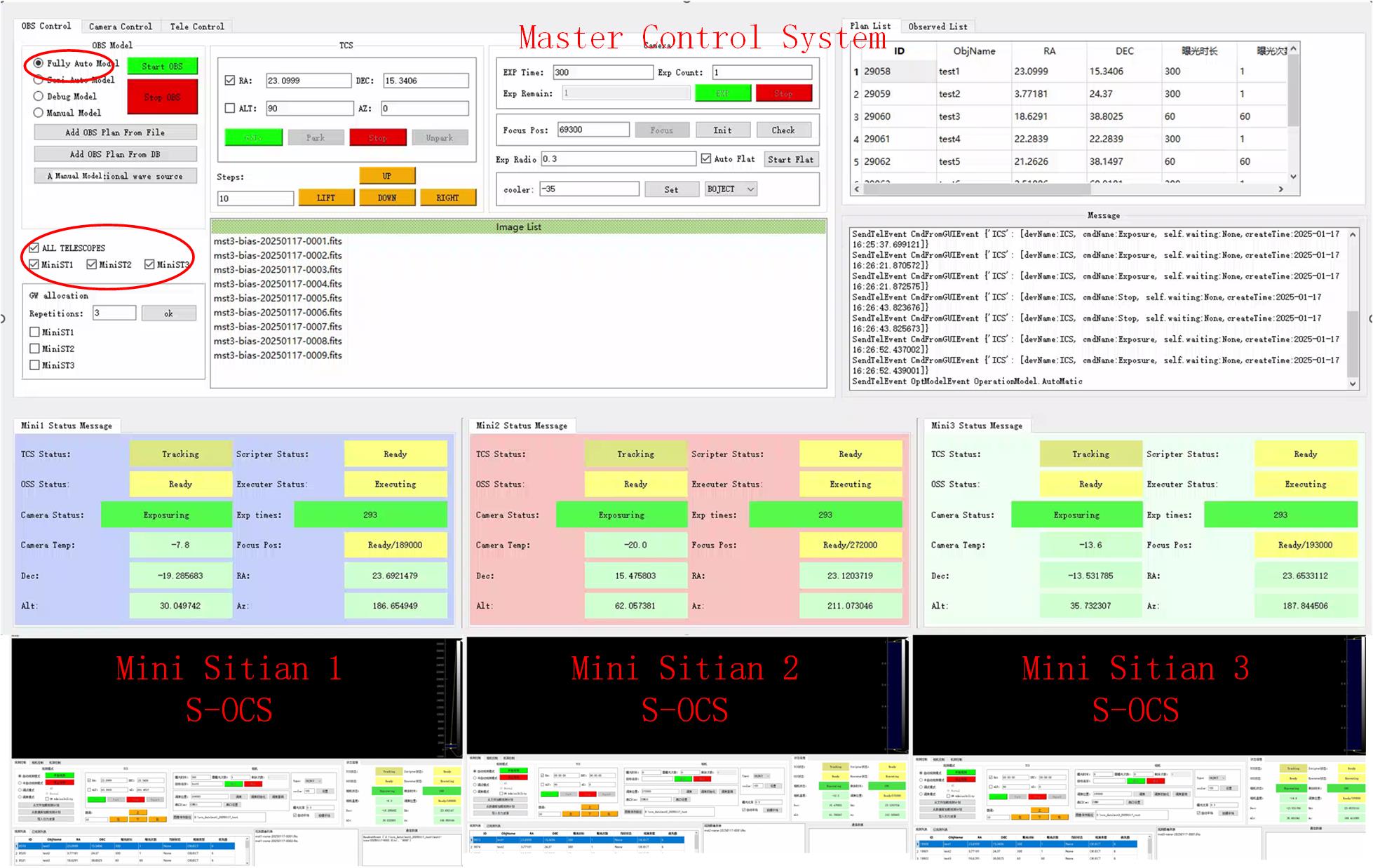}
   \caption{ MCS Auto Model Test}
   \label{Figure4}
   \end{figure}
    
    \item Manual Observation Mode: In this mode, the MCS dispatched a single observation target to a designated telescope or a specified group of telescopes. Upon completion of the observation, an interactive dialog box prompted the operator to manually select the next target. As shown in Figure 5, we selected 'ST3' from the plan list to initiate the observation process. The MCS exclusively sent the target to 'ST3', which began observation. Throughout this process, the remaining two telescopes maintained their initial state and did not execute any commands, ensuring focused and targeted observation as per the operator's discretion.

   \begin{figure}
   \centering
   \includegraphics[width=\textwidth, angle=0]{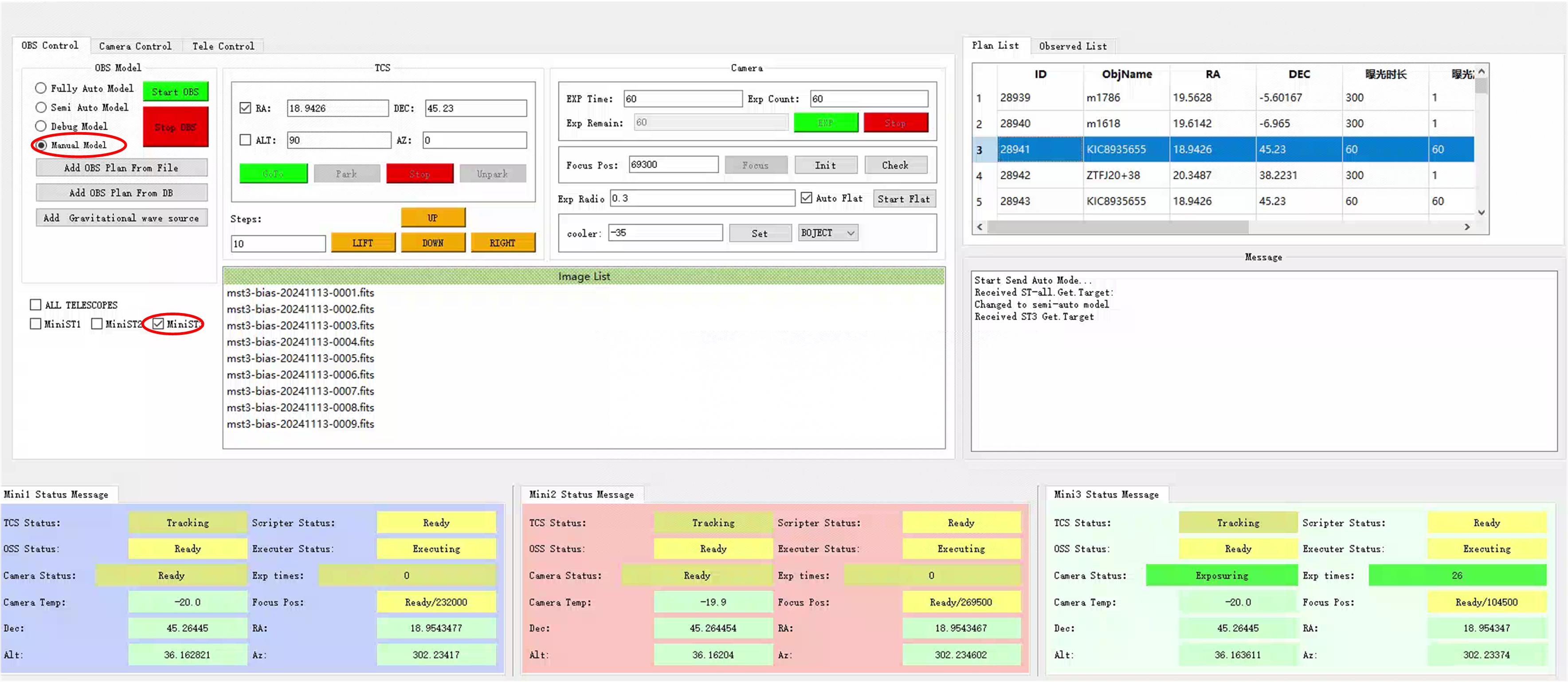}
   \caption{ MCS manual Model Test}
   \label{Figure5}
   \end{figure}
    \item Debugging Mode: In debugging mode, the MCS transmitted specific control commands to the devices of either a single telescope or multiple telescopes, while monitoring the command execution process in real time via the status interface. As depicted in Figure 6, we initially selected 'ST1' and 'ST3' and dispatched a 'Park' command to these telescopes. Subsequently, we deselected 'ST1' and 'ST3' and instead selected 'ST2', sending a 'GoTo' command to direct 'ST2' to point toward the target with coordinates {ra:"18.9428", dec:"45.23"}. Following this, we issued an 'Exposure' command with an exposure time of 60 seconds. Throughout the control sequence, only 'ST2' received and executed the commands accurately, demonstrating precise debugging and control capabilities for individual telescopes within the array.

\end{itemize}
   \begin{figure}
   \centering
   \includegraphics[width=\textwidth, angle=0]{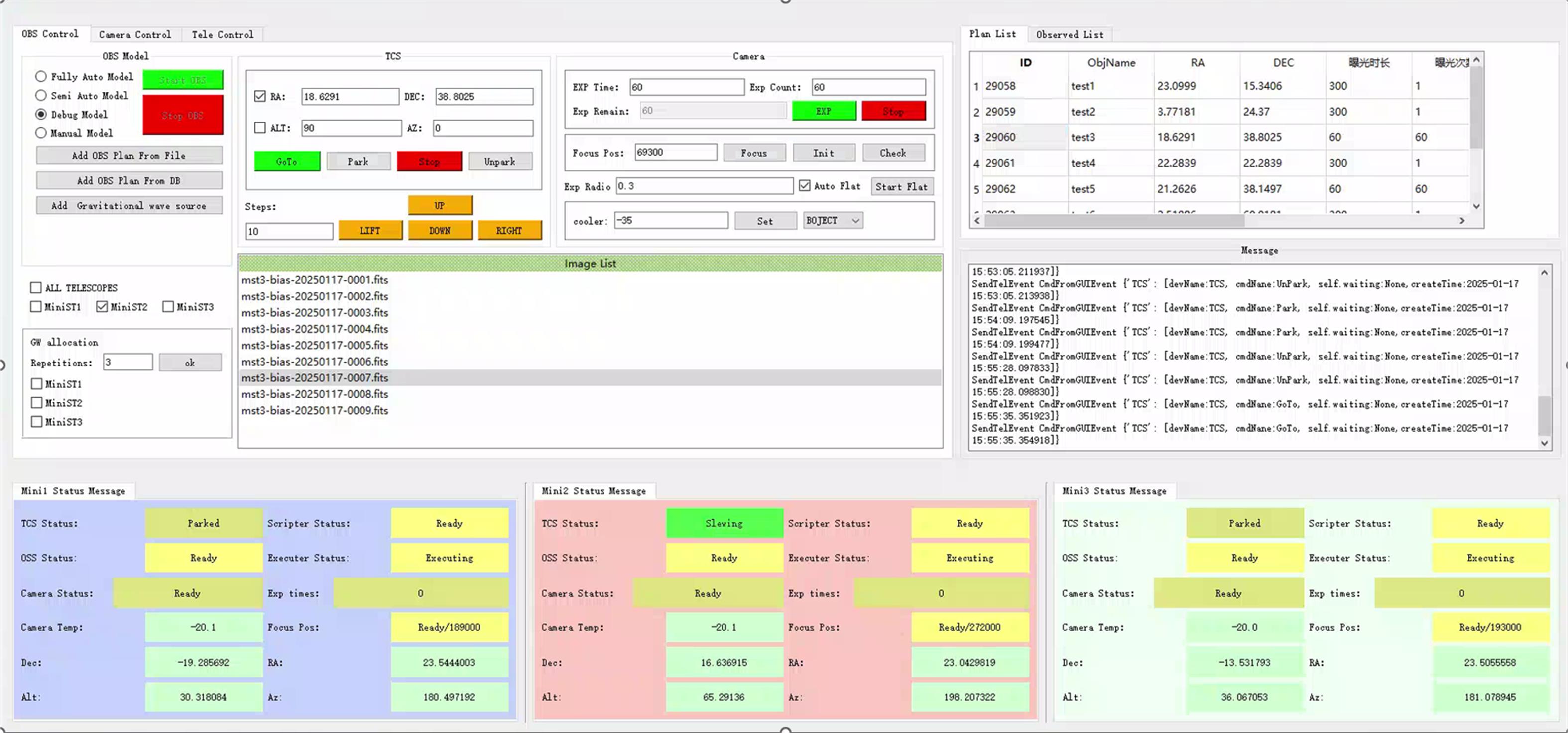}
   \caption{ MCS Debug Model Test}
   \label{Figure6}
   \end{figure}

Through comprehensive testing of various observation modes, the MCS has demonstrated its capability to effectively control the telescope array. In Auto Mode, the MCS enables fully automated control of all telescopes within the array based on the observation list. In Debug Mode, control commands can be selectively sent to specific telescopes without impacting the operation of others. In Manual Mode, the MCS assigns observation targets to one or multiple telescopes, with the corresponding S-OCS executing the automated observation process. These capabilities underscore the MCS's adaptability and efficiency in handling diverse observation scenarios, making it a powerful tool for modern astronomical research.
\subsection{Transient source Testing}
To address special scenarios such as telescope maintenance, the MCS interface includes an option to designate available telescopes. Transient source targets are exclusively allocated to these selected telescopes. Additionally, the MCS interface allows users to specify the desired number of observations for transient source alert targets, providing flexibility in managing observation priorities.

Each telescope's S-OCS receives transient source alerts from the MCS. The S-OCS features a setting to enable or disable the reception of transient source alerts. When enabled, the S-OCS automatically initiates observations of transient source alert targets. If disabled, the S-OCS logs the alert information to a file without interrupting ongoing tasks. During telescope debugging, this option can be deactivated to prevent automatic switching to transient source observation, ensuring uninterrupted debugging processes. Upon receiving a transient source alert, the S-OCS displays a prompt window to notify the observation assistant, enabling timely decision-making.

   \begin{figure}
   \centering
   \includegraphics[width=\textwidth, angle=0]{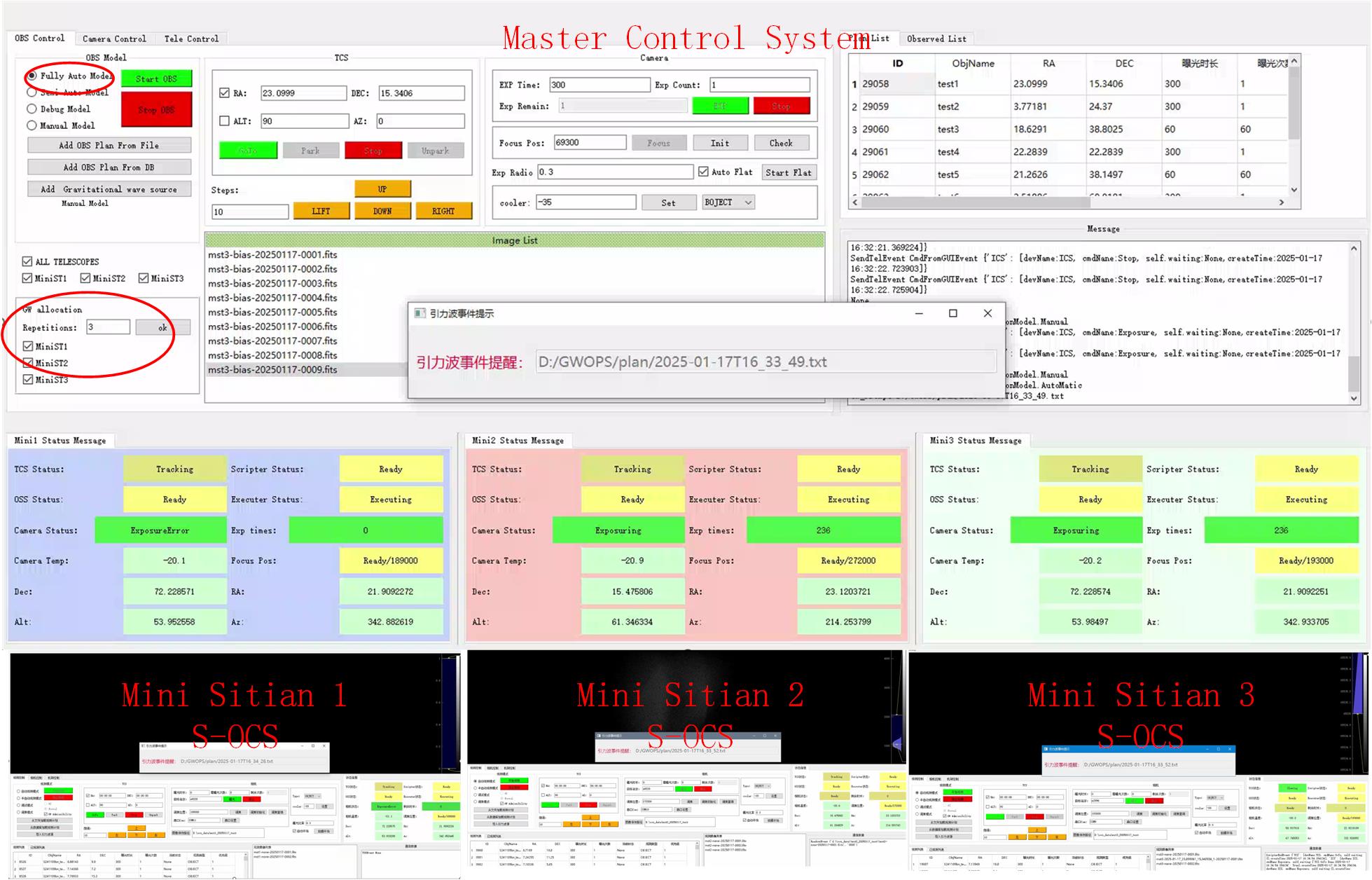}
   \caption{ transient source Test}
   \label{Figure7}
   \end{figure}

The testing of observation plan selection strategies primarily focused on the S-OCS's performance in Fully Automatic Observation Control Mode. As illustrated in Figure 7, the three S-OCSs and the MCS operate in this mode. When transient source alert information is received, it is transmitted to the MCS in the form of a text file. Upon receiving the alert, the MCS triggers a pop-up dialog box to notify the observer of the unexpected event. Simultaneously, the MCS backend extracts the transient source observation target queue and automatically allocates the observation list based on the current pointing positions of the telescopes within the array. This ensures that all targets associated with the current event are observed in the shortest possible time. The observation list is then transmitted to the three telescopes for immediate execution.

After completing the observation of all targets associated with the transient source, the MCS and S-OCS automatically transition to observing the routine observation list. The system then enters a standby state, awaiting the arrival of the next transient source event. This seamless transition ensures continuous and efficient utilization of the telescope array, maintaining optimal observation efficiency.

Through extensive testing and verification, the MCS has demonstrated its ability to efficiently meet diverse control requirements of the telescope array while enabling real-time observation of transient sources. This functionality significantly enhances the observational efficiency of the array and provides robust support for rapid response and dynamic adjustments in astronomical observations. By leveraging automated and intelligent control mechanisms, the MCS can swiftly allocate observation tasks during transient source events, ensuring that scientific targets are captured and analyzed in the shortest possible time. This maximizes the scientific potential of the telescope array and accelerates the discovery of critical astronomical phenomena.

Furthermore, the MCS's multi-mode control capabilities—including Auto Mode, Debug Mode, and Manual Mode—enhance the system's flexibility and adaptability. These modes enable the MCS to address a wide range of observation scenarios and research needs, from routine surveys to targeted observations and system maintenance. The integration of these features ensures that the MCS remains a versatile and powerful tool for modern astronomical research.
\section{Discussion}


Currently, the Control system of the Mini-SiTian Telescope Array comprises multiple S-OCSs and the MCS. While the S-OCSs have successfully achieved single-telescope control in previous observational operations, the MCS, as a newly introduced module, enables coordinated control of the entire telescope array. Additionally, the MCS serves as a critical interface for interaction between the OCS and the SiTian Brain. Looking ahead, the MCS will focus on the development of the following advanced features:
\begin{enumerate}
    \item Meteorological Prediction and Adaptive Observation: Collaborating with meteorological experts, the MCS will integrate meteorological prediction, cloud chart recognition, and dynamic forecasting capabilities. This will allow the observation system to automatically avoid weather-affected areas, ensuring that observations are conducted under optimal conditions. 
    \item Health Status Monitoring and Fault Prediction: The MCS will implement anomaly detection and fault prediction mechanisms for the telescope array. By dynamically adjusting observation plans based on the real-time health status of the telescopes, the system will maximize observation efficiency and minimize potential downtime.
    \item Centralized Deployment and Management: The MCS will promote centralized deployment and management of the telescope array observation control system. This will simplify operational processes, enhance system stability, and improve maintainability, ensuring seamless and efficient operations.
\end{enumerate}
Through these technological upgrades and functional expansions, the MCS is expected to better adapt to complex astronomical observation requirements. It will provide astronomers with more powerful, flexible, and reliable observation control tools, driving advancements in astronomical observation technology and supporting cutting-edge astronomical research.

\section{Summary}
The successful implementation of the MCS in the observation control of the Mini-SiTian array represents a significant milestone in modern astronomical technology. The system is capable of simultaneously receiving and displaying the operational status information of all devices across three telescopes, while also enabling automated control of the entire array through an intuitive user interface. This functionality has greatly simplified the work of operations, allowing them to focus more on data analysis rather than the intricate details of equipment operation.

Beyond equipment monitoring and control, the MCS demonstrates advanced intelligent task management capabilities. By receiving and parsing observation plans from the simulated SiTian Brain, the MCS can intelligently schedule telescopes for automated observations based on task priority and scientific importance. Notably, the system prioritizes high-importance observation tasks, such as gravitational wave events and other critical scientific targets, ensuring that scientists obtain key data in a timely manner to advance astronomical research.

In addition to automated task scheduling, the MCS exhibits remarkable flexibility, adapting to diverse scientific needs and observation conditions. It supports various observation modes, including continuous and periodic observations, which are essential for improving observation efficiency and data quality.

In summary, the successful application of the MCS not only provides a powerful tool for astronomical observation but also injects new momentum into the exploration of the universe. As technology advances and the system undergoes further optimization, we anticipate that the MCS will play an even more pivotal role in the field of astronomical observation, driving the in-depth development of astronomical research.

\begin{acknowledgements}
The SiTian project is a next-generation, large-scale time-domain survey designed to build an array of over 60 optical telescopes, primarily located at observatory sites in China. This array will enable single-exposure observations of the entire northern hemisphere night sky with a cadence of only 30-minute, capturing true color (gri) time-series data down to about 21 mag. This project is proposed and led by the National Astronomical Observatories, Chinese Academy of Sciences (NAOC). As the pathfinder for the SiTian project, the Mini-SiTian project utilizes an array of three 30 cm telescopes to simulate a single node of the full SiTian array. The Mini-SiTian has begun its survey since November 2022. The SiTian and Mini-SiTian have been supported from the Strategic Pioneer Program of the Astronomy Large-Scale Scientific Facility, Chinese Academy of Sciences and the Science and Education Integration Funding of University of Chinese Academy of Sciences.

This work is Supported by National Key R\&D Program of China (Grant No.2023YFA1608304), Strategic Priority Research Program of the Chinese Academy of Sciences (XDB0550103) and National Natural Science Foundation of China (grant No. 11903054, No. 12422303 and No. 12261141690).
\end{acknowledgements}




\bibliographystyle{raa}
\bibliography{ocs}

\begin{thebibliography}{5}
\providecommand\natexlab[1]{#1}
\providecommand\JournalTitle[1]{#1}

\bibitem[Cui {et~al.}(2012)]{cui2012}
Cui, X.-Q., Zhao, Y.-H., Chu, Y.-Q., {et~al.} 2012, Research in Astronomy and
  Astrophysics, 12, 1197

\bibitem[Dworak {et~al.}(2012)]{dworak2010}
Dworak, A., Ehm, F., Charrue, P., {et~al.} 2012, Journal of Physics Conference
  Series, 396, 012017

\bibitem[Kub{\'a}nek {et~al.}(2008)]{peter2008}
Kub{\'a}nek, P., Jel{\'{\i}}nek, M., French, J., {et~al.} 2008, Proceedings of
  SPIE, 7019, 70192S

\bibitem[Kub{\'a}nek {et~al.}(2006)]{peter2006}
Kub{\'a}nek, P., Jel{\'{\i}}nek, M., V{\'{\i}}tek, S., {et~al.} 2006,
  Proceedings of SPIE, 6274, 62741V

\bibitem[Wang {et~al.}(2021)]{wang2021}
Wang, Z., Tian, Y., Li, J., {et~al.} 2021, Research in Astronomy and
  Astrophysics, 21, 149

\end{thebibliography}

\label{lastpage}

\end{document}